# StainNet: A Fast and Robust Stain Normalization Network

**Hongtao Kang[1, 2], Die Luo[1, 2], Weihua Feng[1, 2], Junbo Hu[3,*], Shaoqun Zeng[1, 2], Tingwei Quan[1, 2], And Xiuli Liu[1, 2,*]**

[1]Britton Chance Center for Biomedical Photonics, Wuhan National Laboratory for Optoelectronics-Huazhong University of Science and Technology, Wuhan, Hubei 430074, China

[2]MOE Key Laboratory for Biomedical Photonics, Collaborative Innovation Center for Biomedical Engineering, School of Engineering Sciences, Huazhong University of Science and Technology, Wuhan, Hubei 430074, China

[3]Department of Pathology, Hubei Maternal and Child Health Hospital, Wuhan, Hubei 430072, China

**\* Correspondence:**
Junbo Hu (cqjbhu@163.com), Xiuli Liu (xlliu@mail.hust.edu.cn)

## Abstract

Stain normalization often refers to transferring the color distribution of the source image to that of the target image and has been widely used in biomedical image analysis. The conventional stain normalization is regarded as constructing a pixel-by-pixel color mapping model, which only depends on one reference image, and can not accurately achieve the style transformation between image datasets. In principle, this style transformation can be well solved by the deep learning-based methods due to its complicated network structure, whereas, its complicated structure results in the low computational efficiency and artifacts in the style transformation, which has restricted the practical application. Here, we use distillation learning to reduce the complexity of deep learning methods and a fast and robust network called StainNet to learn the color mapping between the source image and target image. StainNet can learn the color mapping relationship from a whole dataset and adjust the color value in a pixel-to-pixel manner. The pixel-to-pixel manner restricts the network size and avoids artifacts in the style transformation. The results on the cytopathology and histopathology datasets show that StainNet can achieve comparable performance to the deep learning-based methods. Computation results demonstrate StainNet is more than 40 times faster than StainGAN and can normalize a 100,000x100,000 whole slide image in 40 seconds.

## 1    Introduction

Tissues or cells are often transparent and need to be stained before they can be observed under a microscope. However, inconsistencies in the stain reagents, staining process, and scanner specifications often result in different appearances of pathological images [1]. These variations affect the judgment of pathologists, weak the performance of CAD systems, and hamper their applications in pathology [2-4]. Currently, stain normalization algorithms have been proposed to reduce color variation in pathological images. Usually, they transfer the color style of the source image to that of a target image [5] while preserve the other information in the processed image [6]. It is reported that stain normalization helps to increase the prediction accuracy, such as tumor classification, so it has been an important preprocessing task, especially for CAD systems [7]. Stain normalization methods can be broadly classified into two classes: conventional methods and deep learning-based methods.

Conventional methods include Color matching and Stain-separation methods. Color matching methods try to match the color distribution of the source image to that of a reference image [8]. For example, Reinhard et al. [9] calculated the mean and standard deviations of source images and matched them to a reference image in the Lab color space. Stain-separation methods try to separate and normalize each staining channel independently. For instance, Ruifrok and Johnston [10] proposed to measure the relative proportion for three channels (R, G, and B) with the slides stained by only single stain reagent (Hematoxylin or Eosin only) to obtain stain vectors. On the other hand, Macenko et al. [11], Vahadane et al. [12], and Khan et al. [5] used mathematical methods to compute stain vectors. Macenko et al. [11] found the stain vectors by singular value decomposition (SVD) in Optical Density (OD) space. And Vahadane et al. [12] applied sparse non-negative matrix factorization (SNMF) to compute the stain vectors. Khan et al. [5] used a pertained classifier to estimate the relative intensity of the two stains (Hematoxylin and Eosin) to obtain an estimate of the stain vectors. However, Pap stain used in cervical cytopathology involves not only Hematoxylin and Eosin but also Orange, Light Green, and Bismarck Brown [13], which is the main reason why conventional algorithms do not perform well on cervical cytopathology. Nevertheless, most of these rely on a reference image to estimate stain parameters, but it's hard for one reference image to cover all staining phenomena or represent all input images, which usually causes misestimation of stain parameters and thus delivers inaccurate normalization results [14,15].

Deep learning-based methods mostly use generative adversarial networks (GANs) to achieve stain normalization [3,6,8,16-18]. Shaban et al. [8] proposed an unsupervised stain normalization method named StainGAN based on CycleGAN [16]. Furthermore, Cai et al. [3] proposed a new generator, which obtained a better image quality and accelerated the networks. On the other hand, Shaojin et al. [18], Salehi et al. [6] and Tellez et al. [17] reconstructed original images from the images with color augmentations applied, e.g. grayscale and Hue-Saturation-Value (HSV) transformation, and tried to normalize all other color styles to original. However, due to the complexity of deep neural networks and the instability of GANs, it is hard to preserve the source information, and it has a risk to introduce some artifacts, which have some adverse effects for pathological diagnosis [19]. Nevertheless, the network of deep learning-based methods usually contains millions of parameters, so it has low efficiency in computation [14].

Deep learning-based methods perform well in stain normalization, but they are not satisfactory in the robustness and computational efficiency. In this paper, we propose a novel stain normalization network named StainNet, which is a fully 1x1 convolution network to adjust the color value in a pixel-by-pixel manner. And, we use StainGAN [8] as the teacher network and StainNet as the student network to learn the color mapping by distillation learning. The results show that StainNet can achieve comparable performance with StainGAN [8] and better preserved the source information. The results also demonstrate that StainNet was more than 40 times faster than StainGAN [8], which allows StainNet to normalize a 100,000x100,000 whole slide image in 40 seconds.

## 2    Material and Methods

### 2.1    Dataset

In this paper, there are four datasets used to evaluate the performance of different methods. Among them, the aligned cytopathology dataset and the aligned histopathology dataset are used to evaluate the similarity between the normalized image and the target image. The cytopathology classification dataset and the histopathology classification dataset are used to verify normalization algorithms in the classification task. The related descriptions are below:

### 2.1.1 The aligned cytopathology dataset for evaluating the similarity

In this dataset, two slide scanners were used to scan the cervical cytopathology slides from the Maternal and Child Hospital of Hubei Province. One scanner is custom-constructed, called Scanner O, equipped with a 20x objective lens with the pixel size of 0.2930 μm. The other from Shenzhen Shengqiang Technology Co., Ltd., called scanner T, has a 40x objective lens and the pixel size of 0.1803 μm. We resampled the images from scanner T to reduce the pixel size to 0.2930 μm, and then performed rigid and no-rigid registration to align the resampled images to these from scanner O. Finally, 3223 aligned image pairs with the size of 512x512 pixels were collected. Among these images, 2257 pairs of images were randomly selected as the training set, and the remaining 966 pairs of images were used as the test set. The images from the scanner O and T are seen as source images and target images respectively.

### 2.1.2 The cytopathology classification dataset for verifying normalization algorithms

In this dataset, the same data source was used as that in Section 2.1.1, and the patches from scanner T are used as the training set to train the classifier and these from scanner O are used as the test set to evaluate the classifier. In this dataset, we labeled the patches with abnormal cells marked by experts as abnormal patches, and the patches without abnormal cells as normal patches. There are 6589 abnormal patches, 6589 normal patches in the training dataset, 3343 abnormal patches, and 3192 normal patches in the test dataset. The resolution of patches was resampled to 0.4862 um per pixel with dimensions of 256x256. We use StainGAN [8] and StainNet trained on the aligned cytopathology dataset in Section 2.1.1 to normalize the patches in the test set to the style of the training set. Then, we use the original test set and the normalized test set to verify the necessity of stain normalization and evaluate the performance of StainGAN and StainNet.

### 2.1.3 The aligned histopathology dataset for evaluating the similarity

The aligned histopathology dataset is from the publicly available part of the MITOS-ATYPIA ICPR'14 challenge [20]. In the MITOS-ATYPIA dataset [20], there are 16 slides with standard hematoxylin and eosin (H&E) dyes in total, 11 slides used as the training set, and 5 slides used as the test set. And all the slides of this dataset have been scanned by two slide scanners: Aperio Scanscope XT called scanner A and Hamamatsu Nanozoomer 2.0-HT called scanner H. The number of image frames is variable from slide to slide. The training data set contains 1200 frames, and the test data set contains 496 frames at x40 magnification. The resolution of the frames from scanner H was resampled to that of frames from scanner A, and then perform rigid and no-rigid registration to align the resampled frames to these from scanner A. We cropped 16 patches with the size of 256x256 from every frame without overlap, so there are 19200 patch pairs in our training set and 7936 patch pairs in our testing set. In this dataset, the images from the scanner A and H are seen as source images and target images respectively.

### 2.1.4 The histopathology classification dataset for verifying normalization algorithms

The publicly available Camelyon16 dataset [21] is used in this dataset, which contains 399 whole slide images (WSIs) from two centers. In our experiments, the 170 WSIs from Radboud University Medical Center in Camelyon16 training part were used to extract the training patches, and the 50 WSIs from University Medical Center Utrecht in Camelyon16 testing part were used to extract the test patches. We labeled the patches containing tumor cells as abnormal and the patches not containing any tumor cells as normal. For abnormal patches, we extracted patches of size 256x256 from the tumor area in tumor slides. For normal patches, we randomly extracted patches of size 256x256 from the normal area in tumor slides and normal slides until the number of normal patches was equal to the number of abnormal patches. In this way, there are 40K patches in our training set and 10K patches in testing set. In addition, we also randomly extracted 6000 patches from the training set and test set to train StainGAN and StainNet, where the patches from the test set were used as the source image, and the patches from the training set were used as the target image. For the classifier trained on the training set, we used the original test set and the normalized test set to evaluate the classifier, so as to verify the necessity of stain normalization and evaluate the performance of StainGAN and StainNet.

## 2.2 StainNet for stain normalization

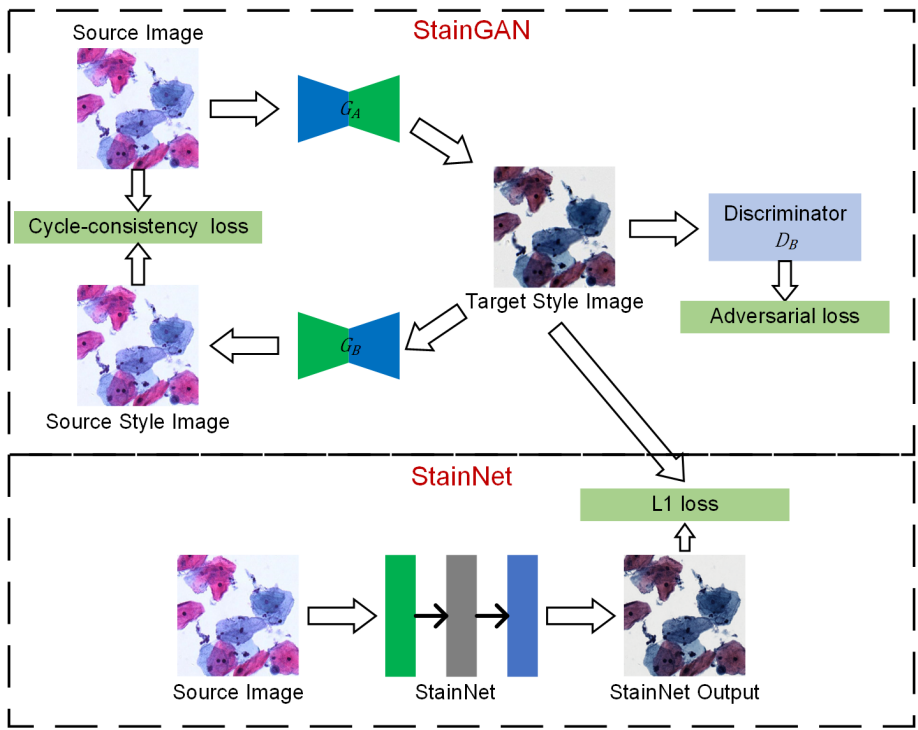

Fig. 1. The framework of StainNet. First, StainGAN normalizes the images from the source domain to the target domain. Then, the normalized images by StainGAN are set as Ground Truth to train StainNet. The images from source domain are mapped to the source domain and then back to the target domain by StainGAN. The same reverse process is also performed for images from the target domain. StainNet is a fully 1x1convolutional neural network, which can directly map the images from the source domain to the target domain.

The framework mainly consists of two parts, one is StainGAN [8], a generative confrontation network with two generators and two discriminators, and the other is StainNet, which is composed of a fully convolutional neural network. StainNet needs paired source and target images to learn the transformation from the source color space to the target color space. It is hard to get the paired images and align the images perfectly. To learn the color mapping relationship, we use StainGAN as the teacher network, and StainNet as the student network, and use the L1 loss to let StainNet learn the output of StainGAN [8].

There are two generators ($G_A$ and $G_B$) and a discriminator ($D_A$ and $D_B$) in StainGAN [8]. $G_A$ is used to transfer the image from source domain to target domain, and $G_B$ is used to transfer from the target domain to source domain. $D_A$ is used to distinguish the image generated by $G_A$ and a real target image, $D_B$ is used to distinguish the image generated by $G_B$ or a real source image. There are two losses in StainGAN, namely cycle-consistency loss, adversarial loss. The cycle-consistency loss [16] ensures that the generated images by $G_A$ can be reconstructed to source image by $G_B$ and the generated images by $G_B$ can be reconstructed to target image by $G_A$. The adversarial loss tries to ensure that the stain distribution of the generated images is consistent with the real distribution.

In the current convolutional neural network, a large number of convolution operations with a kernel size of 3x3 or larger are used. However, a 3x3 or larger convolution performs a weighted summation in the local neighborhood of the input image. Therefore, the pixel value in the output image is inevitably affected by the local neighborhood of the input image. Unlike the 3x3 convolution, the 1x1 convolution only maps a single pixel and has nothing to do with the local neighborhood values. That is, it will not be affected by the texture and can keep the information of inputs. Following this, a fully 1x1 convolutional neural network named StainNet is used to extract the mapping relationship from StainGAN [8]. Except for the last convolutional layer, after each convolutional layer, ReLU is used as a convolutional layer to enhance the nonlinear mapping ability. Considering the balance of performance and computational efficiency, we use three convolutional layers with 32 channels by default. Therefore, our network only contains about 1k parameters, whereas the generator in StainGAN [8] contains millions of parameters.

The training process mainly consists of three steps. Firstly, we train StainGAN [8] using unpaired source and target images. Then, the generator of StainGAN [8] is used to normalize the source images. At last, the normalized images are taken as the Ground Truths to train StainNet with L1 Loss and SGD optimizer. The mapping relationship of StainGAN [8] is based on the image content, that is, the mapping relationship will change accordingly with the different image content. By learning the normalized images by StainGAN [8], StainNet can transfer StainGAN's mapping relationship based on image content into a mapping relationship based on pixel values.

## 3    Experiments and Results

In this section, StainNet is compared with the state-of-the-art methods of Reinhard [9], Macenko [11], Vahadane [12], and StainGAN [8] on the cytopathology and histopathology dataset. We report: 1) Quantitative comparison of different methods in the visual appearance, 2) Application results on the cytopathology and histopathology classification task, 3) Quantitative comparison between the whole slide images normalization results.

### 3.1    Evaluation Metrics

In order to evaluate the performance of different methods, we measure the similarity between the normalized image and the target image, and the consistency between the normalized image and the source image.

Specifically, two similarity metrics are used: Structural Similarity index (SSIM) [21], Peak Signal-to-Noise Ratio (PSNR). The SSIM and PSNR of the target image (SSIM Target and PSNR Target) are used to evaluate the similarity between the normalized image and the target image. The extent of source information preservation is weighed by the SSIM of the source image (SSIM Source), which also was used to measure the similarity between the normalized image and the source image. SSIM Target and PSNR Target are calculated using the original RGB values. SSIM Source is used to measure the preservation of the source image texture information, similar to [22], we use grayscale images to calculate SSIM Source.

Frames per second (FPS) is calculated on the system with 6-core Intel(R) Core (TM) i7-6850K CPU and NVidia GeForce GTX 1080Ti. Input and output (IO) time was not included.

To evaluate the classifier performance, we report the Area Under the Curve (AUC) of the Receiver Operating Characteristics (ROC).

## 3.2 Implementation

For conventional methods Reinhard [9], Macenko [11], and Vahadane [12], a careful picked image was used as the reference image. For the StainGAN, model was trained using Adam optimizer. The training was stopped at 100th epoch, which was chosen experimentally. For StainNet, we first used the trained StainGAN to normalize the source images in both the training dataset and the test dataset. Then we used the normalized images as the Ground Truths during training. StainNet was trained with stochastic gradient descent (SGD) optimizer, an initial learning rate of 0.01, and a batch size of 10. L1 loss was used to minimize the difference between the output of network and the normalized image by the trained StainGAN. Cosine annealing scheduler was adopted to decay learning rate from 0.01 to 0 during 300 epochs. The weights corresponding to the model with the lowest test loss were selected during training.

On the application task, stain normalization was used as a pre-processing step to increase the performance of CAD system. A classifier was trained on the cytopathology classification dataset and histopathology classification dataset to prove this. We used a pretrained SqueezeNet [23] on ImageNet [24] as the classifier and fine-tuned it on the images of the training dataset. The classifier was trained with Adam optimizer, an initial learning rate of 2e-4 and a batch size of 64. Cross-entropy loss was used as our loss function. Cosine annealing scheduler was adopted to decay learning rate from 2e-3 to 0 in 60 epochs. The training was stopped at the 60th epoch, which was chosen experimentally.

## 3.3 Results

In this section, we report the stain transfer results on the aligned cytopathology dataset and aligned histopathology dataset, the classification result on application task, and the WSI normalization results.

### 3.3.1 Stain Transfer Results

First, we evaluate the effectiveness of our method. The normalized images by StainNet are evaluated with the target images through vision and the gray value profiles around the cell nucleus shown in Fig. 2 and Fig. 3. The results on the aligned cytopathology dataset are shown in Fig. 2, where the source images are from scanner O, the target images are from scanner T, and the normalized images in Fig. 2 (c, g) are very close to the target images in Fig. 2 (b, f). The gray value profiles at the nucleus of the source images, target images and normalized images are shown in Fig. 3 (d, h). It's shown that the gray value profiles of the normalized images by StainNet and the target images coincide on the whole, which shows that the color distribution of the normalized images by StainNet is close to that of the target images. In terms of local gray value profiles, the change trend of the normalized images by StainNet is the same as that of the source images, which shows that StainNet can fully retain the information of the source images. The results on the aligned histopathology dataset are shown in Fig. 3, and similar results can also be obtained.

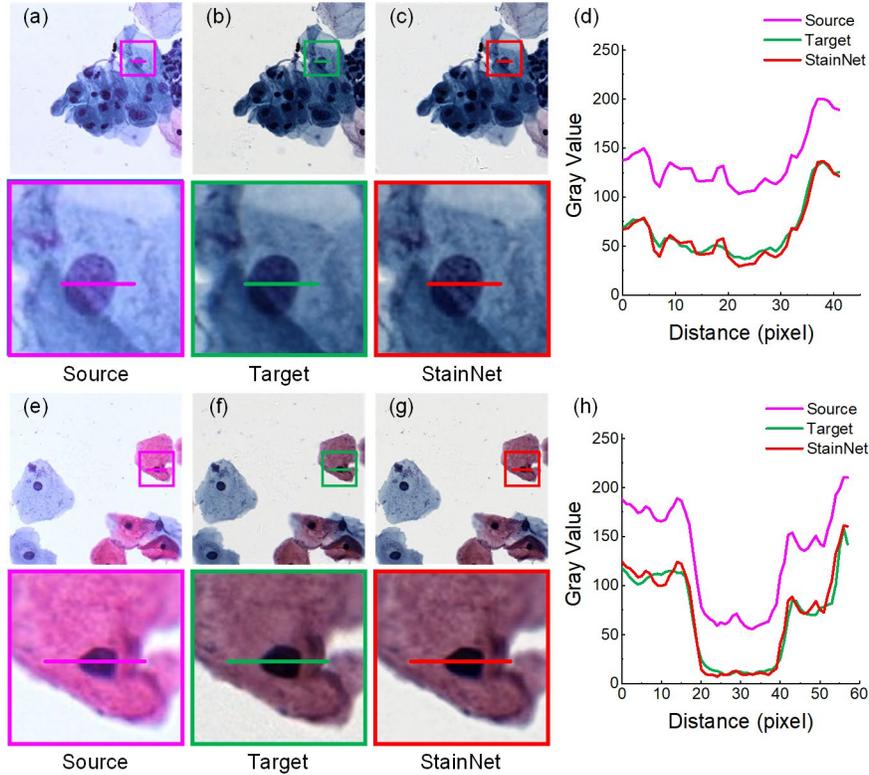

Fig. 2.  StainNet normalized results on the aligned cytopathology dataset. The source images, the target images, and the normalized images by StainNet are shown in (a, e), (b, f), and (c, g), respectively. The image in the dashed box is enlarged below.  Gray value profiles of the straight lines on (a-c) are shown in the line chart (d) and the straight lines in (e-g) are shown in the line chart (h).

Further, StainNet is compared with the other normalization methods. The visual comparison on the aligned cytopathology dataset is shown in Fig. 4. For cytopathological images, the proportion of blank backgrounds is various, so the standard deviation and mean of the different images also are different and we can't find an image to represent the entire dataset. This is the reason that Reinhard method does not perform well in Fig. 4 (c). For Macenko and Vahadane, the color normalization method based on stain separation, it is difficult to perform stain separation correctly due to the use of multiple stains for cytopathological images instead of only eosin and hematoxylin in histopathology. Both StainGAN and StainNet can perform well, but the normalized image of StainNet is more consistent with the source image. The quantitative comparison on the aligned cytopathology dataset is shown in Table 1. In Table 1, the conventional methods are significantly lower than StainGAN and StainNet in PSNR Target, which shows that the conventional methods have a large deviation from the target images, and cannot be directly applied to cytopathological images. StainNet outperform other methods in all indicators. Among them, SSIM Target and SSIM Source are 0.809 and 0.945 higher than StainGAN's 0.764 and 0.905, which shows that StainNet is not only more similar to the target image, but also better to retain the source image information. In addition, the computational efficiency of StainNet is more than 40 times that of StainGAN, which is very important for real-time application.

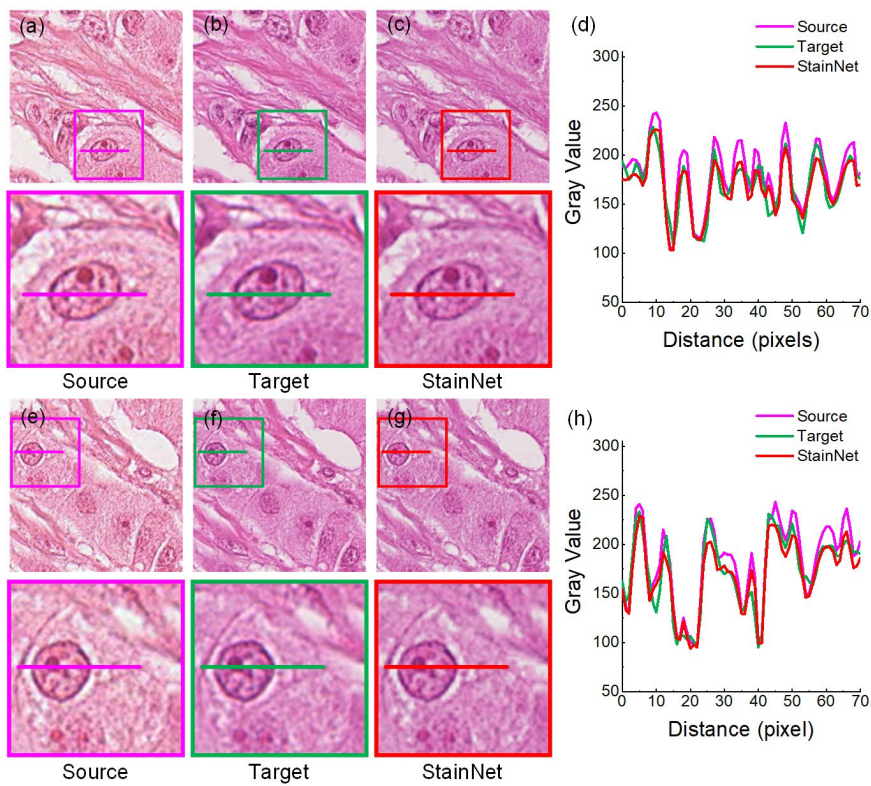

Fig. 3.  StainNet normalized results on the aligned histopathology dataset. The source images, the target images, and the normalized images by StainNet are shown in (a, e), (b, f), and (c, g), respectively. The image in the dashed box is enlarged below.  Gray value profiles of the straight lines on (a-c) are shown in the line chart (d) and the straight lines in (e-g) are shown in the line chart (h).

The visual comparison on the aligned histopathology dataset is shown in Fig. 5. The conventional methods rely on a single reference image to extract the color distribution, which makes it difficult to accurately estimate the color distribution. Therefore, the normalized images by the conventional methods in Fig. 5 (c, d, e) is still visually different from the target image. Although the normalized image can be consistent with the style of the target image, there are obvious texture changes in the normalization result, which is not appropriate in stain normalization. Similar to cytopathological images, the normalized image by StainNet is not only similar to the target image, but also can retain the information of the source image. The quantitative comparison on the aligned histopathology dataset is shown in Table 2. First of all, in this dataset, the test data and training data are completely separated at the slide level and divided in the same way as in the MITOS-ATYPIA ICPR'14 challenge [20], so there is no deviation caused by personal factors. Secondly, due to the rigid and non-rigid registration, the source image and the target image can be precisely matched. The dataset division and image registration make our results more reliable. StainGAN and StainNet are significantly higher than conventional methods in the similarity of SSIM Target and PSNR Target with the target images. The SSIM Target and PSNR Target of StainNet are 0.691 and 22.5 respectively, which are slightly lower than StainGAN's 0.706 and 22.7, but in terms of source image information retention, StainNet's 0.957 is significantly higher than StainGAN's 0.912 in the SSIM Source. Therefore, StainNet can obtain normalized results comparable to StainGAN on the public dataset, but has the ability to better retain the source image information, which may be more important in the real CAD systems.

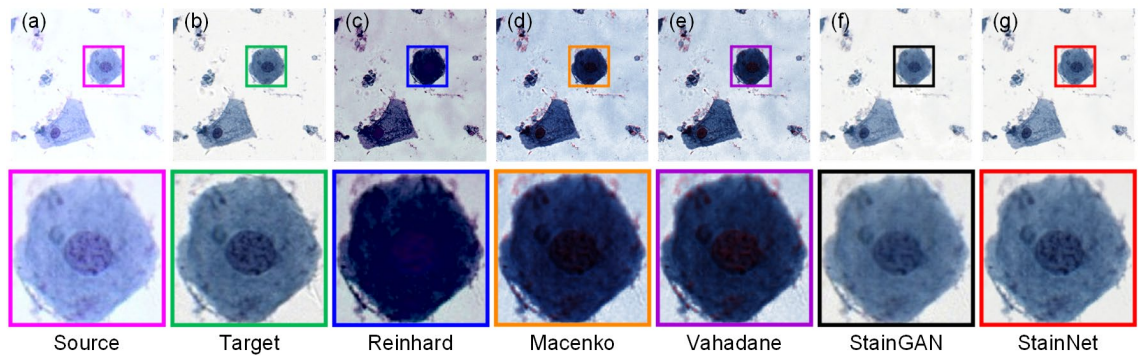

Fig. 4. Visual comparison of different normalization methods on the aligned cytopathology dataset. Source image (a), target image (b), and normalized image by Reinhard (c), Macenko (d), Vahadane (e), StainGAN (f) and StainNet (g) are listed.

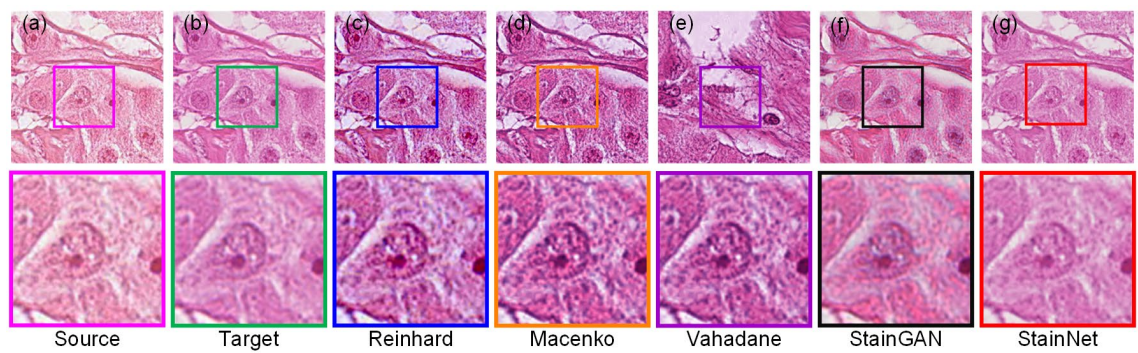

Fig. 5. Visual comparison of different normalization methods on the aligned histopathology dataset. Source image (a), target image (b), and normalized image by Reinhard (c), Macenko (d), Vahadane (e), StainGAN (f) and StainNet (g) are listed.

Table. 1. Different evaluation metrics are reported for various stain normalization methods on the aligned cytopathology dataset.

| Methods | SSIM Target | PSNR Target | SSIM Source | FPS |
|---------|-------------|-------------|-------------|-----|
| Reinhard | 0.739±0.046 | 19.8±3.3 | 0.885±0.042 | 54.8 |
| Macenko | 0.731±0.054 | 22.5±3.1 | 0.853±0.054 | 4.0 |
| Vahadane | 0.739±0.050 | 22.6±3.0 | 0.867±0.050 | 0.5 |
| StainGAN | 0.764±0.030 | 29.7±1.6 | 0.905±0.021 | 19.6 |
| StainNet | 0.809±0.027 | 29.8±1.7 | 0.945±0.025 | 881.8 |

Table. 2. Different evaluation metrics are reported for various stain normalization methods on the aligned histopathology dataset.

| Methods | SSIM Target | PSNR Target | SSIM Source |
|---------|-------------|-------------|-------------|
| Reinhard | 0.617±0.106 | 19.9±2.1 | 0.964±0.031 |
| Macenko | 0.656±0.115 | 20.7±2.7 | 0.966±0.049 |
| Vahadane | 0.664±0.116 | 21.1±2.8 | 0.967±0.046 |
| StainGAN | 0.706±0.099 | 22.7±2.6 | 0.912±0.025 |
| StainNet | 0.691±0.107 | 22.5±3.3 | 0.957±0.007 |

Finally, we compared StainNet and StainGAN in classification application. SqueezeNet [22] pretrained on ImageNet [23] was chosen as the classifier because of its small size and relatively high accuracy. On the cytopathology classification dataset, we use 13,178 image patches from scanner T to train the classifier and use 6,535 image patches from scanner O to evaluate the classifier. On the histopathology classification dataset, the classifier was trained with 40,000 image patches from Radboud University Medical Center, and the classifier was evaluated with 10,000 image patches from University Medical Center Utrecht. For the original images in the test set, there is only an AUC of 0.832 on the cytopathology classification dataset, and only 0.685 on the histopathology classification dataset. It shows that the classifier has a strong color bias, and cannot be directly applied to the test data which have different color style from the training data. The AUC can be increased to 0.896 and 0.905 by using StainGAN, 0.901 and 0.895 by using StainNet on the cytopathology classification dataset and histopathology classification dataset. The above results show that both StainGAN and StainNet can effectively improve the accuracy of the classifier, and the performance of StainNet method and StainGAN method is comparable.

### 3.3.2 Whole Slide Images Results

Table. 3. The SSIM Source of the normalized whole slide image by StainGAN and StainNet.

|  | StainGAN | StainNet |
| --- | --- | --- |
| The cytopathology WSIs | 0.905±0.093 | 0.954±0.050 |
| The histopathology WSIs | 0.762±0.182 | 0.980±0.013 |

For a whole slide image (WSI), there are two mainly challenge in stain normalization. One is that WSIs are very large: a typical WSI may contain 100,000x100,000 pixels. So computational efficiency is very important. The other is that WSIs may contain many naturally-occurring and human-induced artifacts, e.g. air bubbles, dust, and out-of-focus. So the methods must be robust to these phenomena when they are applied in real-world system. Since StainNet has a very concise structure and only maps based on color values, it is less affected by the distribution of the training data and has better robustness.

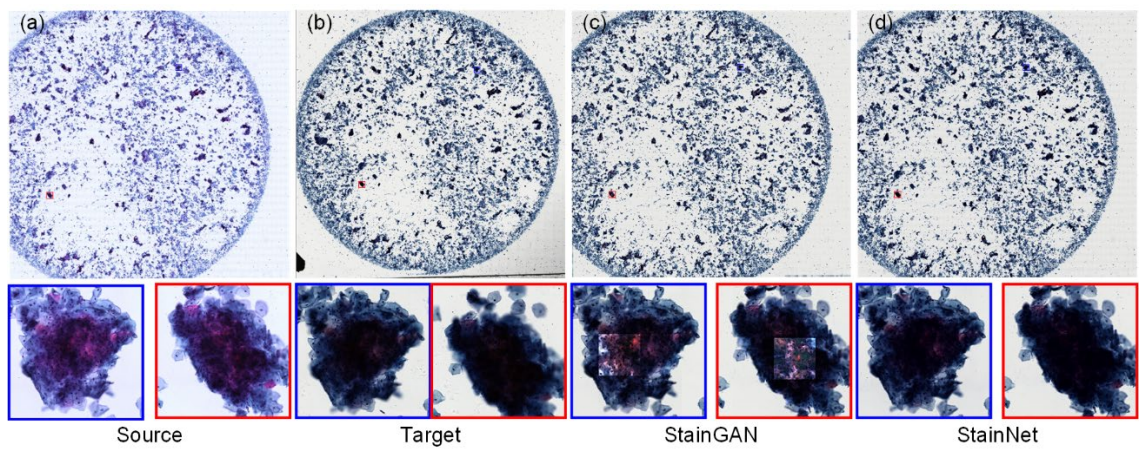

Fig. 6.   The whole slide image normalization result on the cytopathology dataset. Source slide (a), Target slide (b), the normalized slide by StainGAN (c) and the normalized slide by StainNet (d) are list.

In this experiment, we randomly selected 20 cytopathology WSIs from the same data source in Section 2.1.1 and 20 histopathology WSIs from the Camelyon16 dataset. StainGAN and StainNet are used to normalize these WSIs to the target style.

For the cytopathology WSIs, StainGAN and StainNet trained on the aligned cytopathology dataset were used to perform normalization, as shown in Fig. 6. In Fig. 6, the normalized WSI by StainGAN has artifacts in the center of crowded cell clusters. Our proposed StainNet can achieve good results because of its robustness and less relying on the distribution of training set.

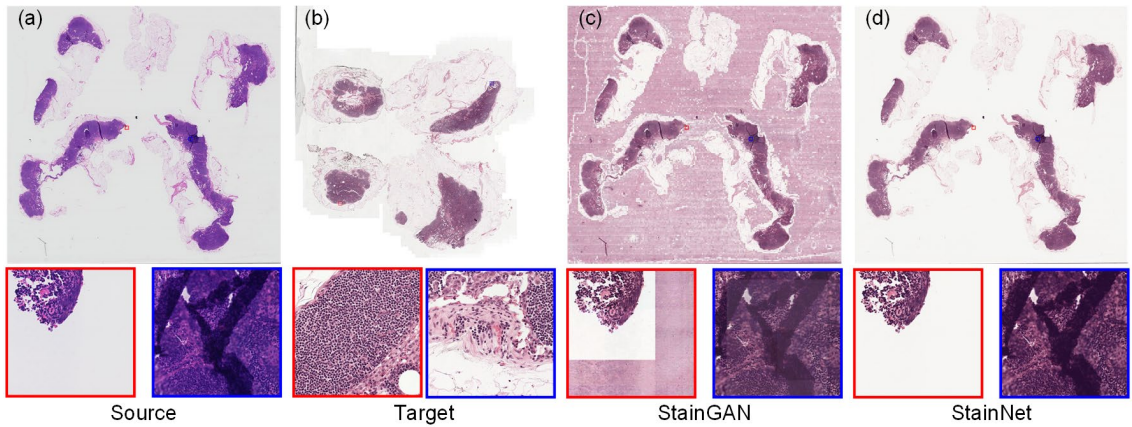

Fig. 7.    The whole slide image normalization result on the Camelyon16 dataset. Source slide (a), Target slide (b), the normalized slide by StainGAN (c) and the normalized slide by StainNet (d) are list.

For the histopathology WSIs, StainGAN and StainNet trained on the histopathology classification dataset were used to perform normalization, as shown in Fig. 7. StainGAN has artifacts in the blank background area and the out-of-focus area. Similar to the cytopathology WSIs, StainNet can still achieve good performance.

For quantitative evaluation, SSIM Source was used to quantitatively compare the normalized performance by StainGAN and StainNet in this experiment. In Table 3, StainNet can obtain a higher mean value and a lower standard deviation, which shows that StainNet not only can obtain better image quality but also has consistent and robust performance on the WSIs. The standard deviation of StainGAN is significantly increased, which shows that the performance of StainGAN is not stable enough on the WSIs.

## 4    Discussion and Conclusion

In this section, we discuss the structure of StainNet to explain the network structure and summarize our work.

In this section, we conduct a comparative experiment to verify the role of 1x1 convolution in stain normalization on the aligned cytopathology dataset. The effectiveness of 1x1 convolution is verified by replacing the three 1x1 convolution in StainNet with 3x3 convolutions in turn. The source image, target image, and normalized image by different structure of StainNet are shown in Fig. 8 (a-f), and the gray value profiles of the straight lines in Fig. 8 (a-f) are shown in Fig. 8 (g). It is clearly that with the increase of 3x3 convolution, the normalized image becomes more blurred, and the ability to preserve the source information is getting worse. The best image quality can be obtained fully using 1x1 convolution in Fig. 8 (c). In particular, at the place pointed by the black arrow in Fig. 8 (g), only a fully 1x1 convolutional network can best preserve the grayscale changes of the source image. The different evaluation metrics, SSIM Target, PSNR Target, and SSIM Source, for different structure of StainNet are reported in Table II. Although the 3x3 convolutions may help improve the similarity with the target images, it affects the ability to preserve the source information. Not changing the information of the source image is a basic requirement for stain normalization, so a fully 1x1 convolutional network is chosen.

Table. 4. Evaluation metrics of different StainNet structures.

| Number of Conv1x1 | Number of Conv3x3 | SSIM Target | PSNR Target | SSIM Source |
|:-:|:-:|:-:|:-:|:-:|
| 3 | 0 | 0.808 | 29.8 | 0.960 |
| 2 | 1 | 0.814 | 30.0 | 0.958 |
| 1 | 2 | 0.814 | 30.0 | 0.956 |
| 0 | 3 | 0.804 | 29.8 | 0.950 |

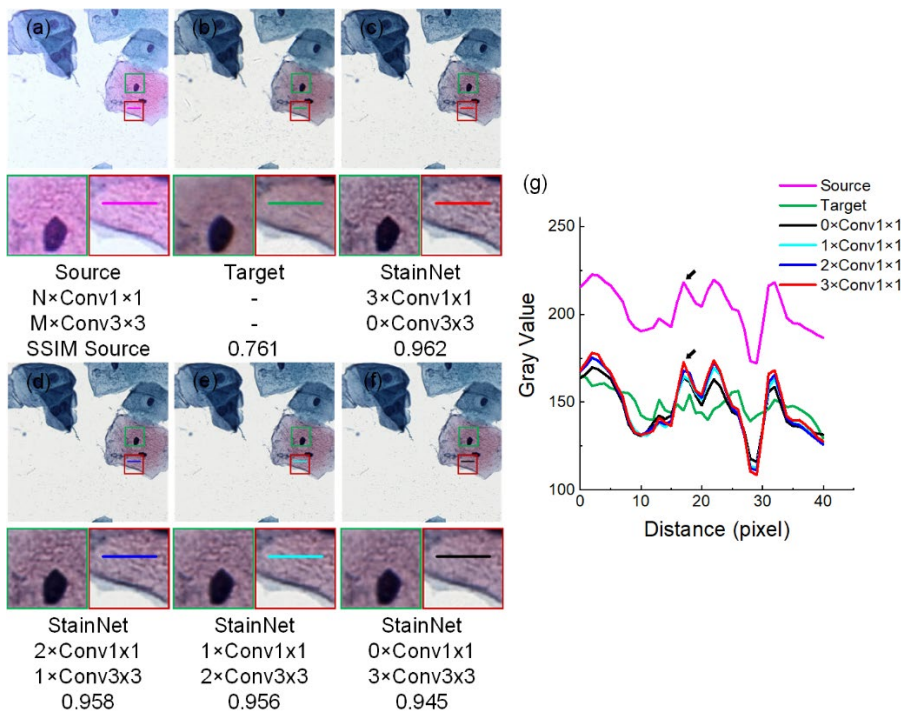

Fig. 8. Effects of 1x1 and 3x3 convolutions. NxConv1x1 and MxConv3x3 refers to the number of 1x1 convolution and 3x3 convolution. StainNet contains only three convolution layers, so the total number of 1x1 convolution and 3x3 convolution is three, that is, M+N=3. The image in the dashed box is enlarged below. Gray value profiles of the straight lines in (a-f) are shown in the line chart (g).

In this paper, we achieve stain normalization by using a fully 1x1 convolutional network in a pixel-to-pixel manner, which not only avoids the low computational efficiency and possible artifacts of deep learning-based methods but also preserves well the information of the source image. Compared with conventional methods, StainNet learns the mapping relationship from the whole dataset instead of relying on one single reference image, so it can obtain the normalized image with high similarity. Furthermore, StainNet has been validated on four datasets including two public datasets, and the results show that StainNet has better performance, especially in computational efficiency and robustness. In short, our work has the potential to perform normalization in real-time in a real-world CAD system.

## 5    Data Availability Statement

The source code of the StainNet network employed in this paper is available at Github: https://github.com/khtao/StainNet.

## 6 Conflict of Interest

The authors declare that the research was conducted in the absence of any commercial or financial relationships that could be construed as a potential conflict of interest.

## 7 Author Contributions

HK contributed to the conception, implemented the experiments and wrote the first draft of the manuscript. JH provided a cytopathology dataset, annotated the cytopathology images. HK, DL, WF, JH, SZ TQ and XL design of the study, and contributed to the result analysis and manuscript revision. All authors approved the final manuscript.

## 8 Funding